# High Reynolds number airfoil turbulence modeling method based on machine learning technique

Xuxiang Sun, Wenbo Cao, Yilang Liu, Linyang Zhu, Weiwei Zhang,

Northwest Polytechnical University, Xi 'an, 710072 China

**Abstract:** In this paper, a turbulence model based on deep neural network is developed for turbulent flow around airfoil at high Reynolds numbers. According to the data got from the Spalart-Allmaras (SA) turbulence model, we build a neural network model that maps flow features to eddy viscosity. The model is then used to replace the SA turbulence model to mutually couple with the CFD solver. We build this suitable data-driven turbulence model mainly from the inputs, outputs features and loss function of the model. A feature selection method based on feature importance is also implemented. The results show that this feature selection method can effectively remove redundant features. The model based on the new input features has better accuracy and stability in mutual coupling with the CFD solver. The force coefficient obtained from solution can match the sample data well. The developed model also shows strong generalization at different inflow condition (angle of attack, Mach number, Reynolds number and airfoil).

## 1. Introduction

Reynolds-averaged Navier-Stokes equations (RANS) remains the primary tool for simulating high-Reynolds number turbulent flows in the aerospace industry[1]. However, many drawbacks limit its application heavily, such as the variability between the results of different turbulence models. Therefore, developing effective and applicable closure models has become an important issue in turbulence modeling work. For this problem, previous studies are mainly based on semi-empirical or semi-theoretical approaches. Parameters of RANS models are most often estimated from canonical situations, which are often "overfitted" and generalize poorly in realistic applications.

Following experiments, theories and computations, data has become the fourth paradigm for people to understand nature[2]. Related data-driven algorithms have made notable progress in many fields including image recognition[3] and natural language processing[4]. In turbulence modeling, some researchers have also started relevant work to seek solutions to traditional problems from the new perspective of data-driven methods. Recent studies have shown promising results on data-driven approaches for turbulence modeling.

Ideas of using data-driven method to study turbulence model can be divided into the three categories[5]. (1) Modeling source terms in turbulence model to improve its performance. Duraisamy et al.[6-9] used a neural network model to model the source terms in the SA model and embed it in the computational fluid dynamics (CFD) solver. (2) Directly modeling Reynolds stress obtained from DNS dataset. For example, Wang, Xiao et al[10,11] modeled discrepancy of Reynolds stress between RANS and DNS method using random forest model. It indicated that the obtained model can improve RANS-predicted Reynolds stress both in training and testing flows. Ling and Templeton et al.[12,13] constructed the model's characteristic quantities in the form of Galileo invariants and predicted the unreasonable eddy viscosity regions computed by traditional models through machine learning. (3) Building a new machine learning model from data to replace the traditional turbulence model. For example, Ling et al.[14] used a tensor basis neural network to model the Reynolds stress. The simulation results were much better than the traditional RANS model. Zhu et al.[15] used a single-layer radial basis function (RBF) network to construct a pure data-driven turbulence model and realized the mutual coupling with the CFD solver. The proposed model can achieve the same accuracy and higher computational efficiency as the SA model based on the three subsonic examples of NACA0012 airfoil. It also has a certain generalization ability for the computational state and geometric shape, which verifies the feasibility of the alternative model. Maulik et al.[16] developed a physics-constrained eddy-viscosity surrogate model for RANS and achieved a 5-7 times acceleration. Zhao et al.[17] presented a novel CFD-driven machine learning framework to develop RANS models. This method is designed to train a model that can well adapt to the average RANS velocity and turbulence scale, so the obtained model can be well coupled with the CFD solver for solution. Beck et al[18,19] used a convolutional residual neural network model to close the stress terms in the large eddy simulation (LES) and coupled it with the CFD solver. Gamahara et al[20] trained a neural network model to predict components of the subgrid stresses in the LES using DNS data and examined the effect of the input features of the model. Nikolaou et al[21] used high-fidelity DNS/LES data to train a neural network to predict six independent components of the stress tensor, and achieved good results on both the training and test samples.

The above research work has shown the feasibility of applying machine learning methods in turbulence modeling. Machine learning has already triggered some changes in this field. And it will play a key role in the modeling of complex flows in the future[22]. The purpose of improving or constructing the turbulence model is to improve the simulation of the flow and to make the final solution of the flow field more similar to the real physical situation. Therefore, coupling the obtained model and NS equation is also an important step. Related work has shown some problems, mainly in the stability and convergence of the coupling process. It is a complex problem and some scholars have studied it from the properties of RANS equation[23,24] and the propagation error[25,26]. In addition to these, the discontinuity and non-physical properties of the data-driven model`s output also lead to coupling failure [18]. The purely data-based modeling does not apply in the physical problems. Combining data and prior knowledge of physics may be a correct way. The a priori knowledge mentioned here is not just partial differential equations, but can also include physical formulas, empirical knowledge, and even data distribution, which can be added to the construction of the model by some means to guide the training process of the model and make the output more physically consistent. Researchers have made some explorations in this regard. Karpatne et al. proposed a physics-guided neural network[27]. They added the physical prior information into loss function to guide the training of model. In this way, neural network not only has a high accuracy, but also the predicted results conform to the physical consistency principle. Raissi et al.[28] proposed the concept of PINNs (Physics-Informed Neural Networks), and introduced general nonlinear partial differential equations as constraints into the training of neural networks. Using this method, data-driven solutions and inverse modeling of partial differential equations can be realized. Based on the idea of PINNs, they realize the goal of predicting the flow field and identifying related parameters[29,30]. PINNs make full use of the universal approximation property of neural networks, but lack of generalization, which will seriously limit its application. Besides, Lutter et al[31] encoded the Lagrangian equations into the neural network to ensure the smoothness and physical rationality of the model`s output.

The past work is mainly applied to turbulence modeling with low Reynolds number and simple shape. It is usually made up or modified based on the traditional partial differential equation. Few of them achieve mutual coupling with NS equations. In our previous research work[15], a single-layer radial basis neural network model and the idea of partition modeling were adopted to initially realize the complete replacement of the traditional turbulence model. Based on it, a deep neural network method is used to construct a full-field unified turbulence model in this paper. We improve the model in terms of the construction and selection of model input features and the design of the loss function We also study the stability and convergence of the mutual coupling with NS equation.

## 2. Methodology

### 2.1 Modeling framework

In previous work, we have modeled the subsonic attached flow around an airfoil with a single-layer radial basis neural network. Due to the limit of single-layer RBF network, we use the strategy of partitioned modeling, which is hard to implement in practice because of its low operability. Moreover, RBF networks are prone to problems such as ill-conditioned when optimizing parameters, which makes it difficult to get optimal values. To avoid these problems, in this paper we choose a deep fully connected neural network (DNN) for modeling. Thanks to the powerful nonlinear expression ability of DNN, we abandon the idea of partition modeling and achieve unified modeling of the entire flow field. The scope of modeling has also been extended to transonic and larger Reynolds numbers. The main modeling framework of this paper is shown in Figure 1. The work is divided into two parts: modeling process and coupling process. The modeling process includes data acquisition, data preprocessing, features construction and selection, and neural network training. The coupling process refers to the process of replacing the original turbulence model with the obtained DNN model, coupling with the CFD solver, participating in the iterative solution of the flow field, and finally obtaining a convergent flow field.

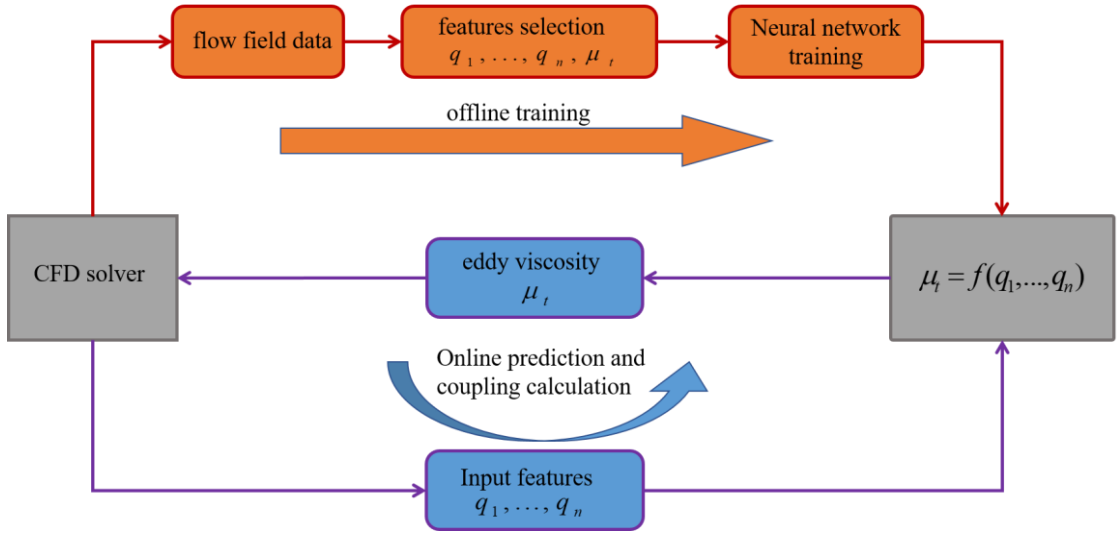

Figure 1　main modeling framework

## 2.2 RANS

In Cartesian coordinates for an incompressible, constant-property fluid, the RANS equations can be written as follows：

$$\frac{\partial U_i}{\partial x_i} = 0$$

$$\rho\frac{\partial U_i}{\partial t} + \rho U_j \frac{\partial U_i}{\partial x_j} = -\frac{\partial P}{\partial x_i} + \frac{\partial}{\partial x_j}\left(2\mu S_{ij} - \rho\overline{u_i' u_j'}\right) \qquad (1)$$

The vectors $U_i$ and $x_i$ are mean velocity and position, $t$ is time, $P$ is mean pressure, $\rho$ is density and $S_{ij}$ is the strain-rate tensor:

$$S_{ij} = \frac{1}{2}\left(\frac{\partial U_i}{\partial x_j} + \frac{\partial U_j}{\partial x_i}\right) \qquad (2)$$

$\overline{u_i'u_j'}$ is a time-averaged rate of momentum transfer due to the turbulence. And The quantity $\rho\overline{u_i'u_j'}$ is known as the Reynolds-stress tensor. In order to compute all mean-flow properties of the turbulent flow under consideration, we need a prescription for computing $\overline{u_i'u_j'}$. In this work, data used for modeling are all generated from the flow around NACA0012 airfoil simulated by Spalart-Allmaras (SA) turbulence model[32]. The transport equation of the SA model is as follows:

$$\mu_t = \rho\hat{v}f_{v1} \tag{3}$$

$$\frac{D\hat{v}}{Dt} = C_{b1}(1-f_{t2})\hat{S}\hat{v} + \frac{1}{\sigma}\left[\frac{\partial}{\partial x_j}\left((v+\hat{v})\frac{\partial \hat{v}}{\partial x_j}\right) + C_{b2}\frac{\partial \hat{v}}{\partial x_i}\frac{\partial \hat{v}}{\partial x_i}\right] - \left(C_{w1}f_w - \frac{C_{b1}}{\kappa^2}f_{t2}\right)\left(\frac{\hat{v}}{d}\right)^2 \tag{4}$$

$$\begin{aligned}
&f_{v1} = \frac{\chi^3}{\chi^3 + c_{v1}^3}, \chi = \frac{\hat{v}}{v}, \hat{S} = \Omega + \frac{\hat{v}}{\kappa^2 d^2}f_{v2}, \Omega = \sqrt{2W_{ij}W_{ij}} \\
&f_{v2} = 1 - \frac{\chi}{1+\chi f_{v1}}, f_w = g\left[\frac{1+c_{w3}^6}{g^6 + c_{w3}^6}\right]^{1/6}, g = r + c_{w2}(r^6 - r), \\
&r = \min\left[\frac{\hat{v}}{\hat{S}\kappa^2 d^2}, 10\right], f_{t2} = c_{t3}\exp\left(-c_{t4}\chi^2\right), W_{ij} = \frac{1}{2}\left(\frac{\partial u_i}{\partial x_j} - \frac{\partial u_j}{\partial x_i}\right), \\
&c_{b1} = 0.1355, \sigma = 2/3, c_{b2} = 0.622, \kappa = 0.41, \\
&c_{w2} = 0.3, c_{w3} = 2, c_{v1} = 7.1, c_{t3} = 1.2, c_{t4} = 0.5, c_{w1} = \frac{c_{b1}}{\kappa^2} + \frac{1+c_{b2}}{\sigma}
\end{aligned} \tag{5}$$

where the turbulent eddy viscosity is defined as $\mu_t$ and the stress tensor is $S$, with $\Omega$ being the rotational tensor, $d$ is the distance from the closest surface.

## 2.3 Modeling strategy

At present, the cost of LES or DNS simulation with high Reynolds number (Re) flow is unacceptable[33]. Here we take the data from the RANS model as the modeling data. The main content of this paper is how to effectively get a usable turbulence model from data . Therefore, data from RANS model meets the requirements. When using higher resolution data, the modeling ideas and methods proposed in this paper also provide an effective reference.

To ensure the sample data covers different flow states as much as possible, we prefer to select boundary points as training samples. Then use the Latin hypercube sampling method to sample the internal angle of attack α and Mach number (Ma). Reynolds number is not sampled here since it is a discrete variable. Instead, a random sampling method is used to assign a Reynolds number to each sampling result. The range of Angle of attack α is $[0°,12°]$, the value range of Ma is $[0,0.8]$, and the range of Reynolds number Re is $\{2\times10^6, 3\times10^6,...,9\times10^6\}$. Under a high Mach number and a high angle of attack, the flow around the airfoil will be instable, which is beyond the scope this paper. Therefore, the sampling area is a trapezoid rather than a rectangle to avoid this(Figure 2).

Sampling data is divided into training set and test set. Then 70% of the training set is used for modeling, and the remaining 30% is used as the validation set to evaluate the model in the modeling process. The test set is used to evaluate the performance of the model under unknown conditions. The final sampling results and the data set division are shown in Table 1.

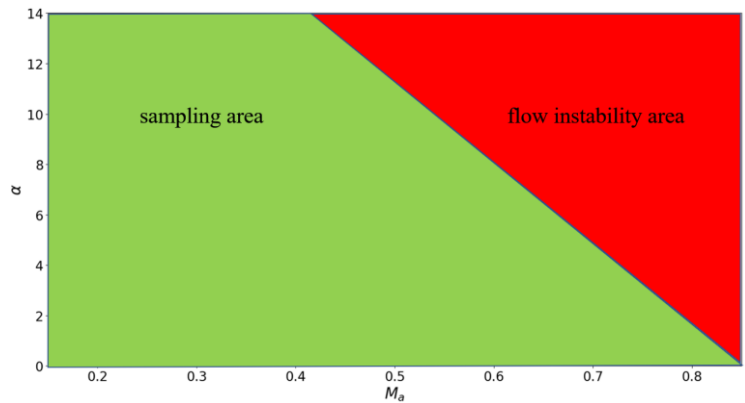

Figure 2 Schematic diagram of sampling area

Table 1 Data division of training set and test set

| Case | α/° | Ma | Re/$10^6$ |
|---|---|---|---|
| | 0 | 0.1 | 2,9 |
| | 0 | 0.7 | 2,9 |
| Training set | 12 | 0.1 | 2,9 |
| | 4.1 | 0.7 | 2,9 |
| | 5.07 | 0.4 | 5,8 |

|             | 1.23 | 0.68 | 6,8 |
|             | 2.48 | 0.71 | 4,7 |
| Testing set | 2.48 | 0.71 | 5   |
|             | 5.07 | 0.4  | 4   |

According to the existing work, there are two main ways to use the flow field data. (1) Modeling based on grid point data. That is, the data on each grid point is used as a separate input term, and this approach applies to fully connected neural networks[15,34] ; (2) Modeling based on full-field data. That is, the whole flow field is used as an input term, and it applies to convolutional neural networks[35,36]. In this paper, we use the first approach since this point-to-point modeling approach can fully consider the specificity of the flow at each grid point. For example, for a transonic flow field, which contains three flow states: sub, trans, and supersonic. The data at this point contains richer information about the flow field. Therefore, it is possible to train the model more effectively while reducing the amount of data. The model obtained in this way reflects only a local mapping of the flow features to the eddy viscosity. As for the problem that the local model is not sensitive to the global flow parameters, it can be solved by data pre-processing and reasonable construction of input and output features.

## 2.4 Feature construction and selection

A machine learning algorithm relies heavily on the input features[37]. Good input features can speed up learning, improve model accuracy, and contribute to the generalization ability of the model. Removing redundant features can reduce the complexity of the model and the difficulty of learning[38].

There are two general types of feature construction: manual construction and automatic construction. Manual construction of features needs personal with certain domain knowledge to select good features through multiple experiments. Automatically constructing features is a hot topic in machine learning today. It belongs to the category of automatic machine learning and aims to replace the complicated feature construction and selection through some advanced algorithms. Several algorithms for automatic feature engineering have emerged and achieved good results, such as FeatureTools[39]. However, these automatic feature construction methods can only apply to some simple problems in daily life. For complex physical problems such as turbulence, these methods cannot provide a good feature set. In many cases, they just traverse and combine the initial features. It is far from building features with clear physical meaning. Therefore, in this paper, we still construct the features manually.

Based on the previous work, we combined our knowledge and experience of the physical problem to construct the initial set of input features shown in Table 2. The reason they are initial features here is because we have selected them later to reduce the number of features without affecting the modeling accuracy. A plain idea of constructing features is to build features similar to the target distribution. For example, according to the flow field of NACA0012 airfoil at $Ma=0.15, \alpha=0°, Re=3\times10^6$, the contour of features and eddy viscosity are shown in Figure 3.

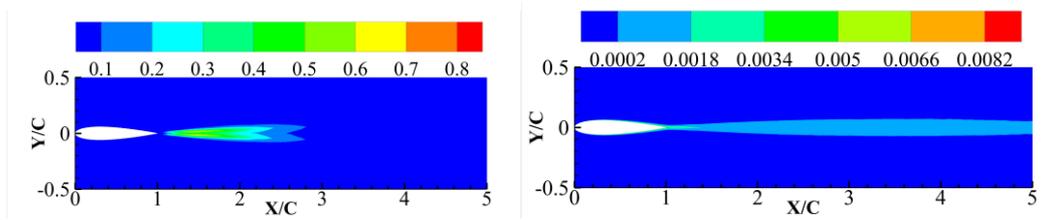

Figure 3 Left: contour of $q_8$; Right: contour of $q_3$

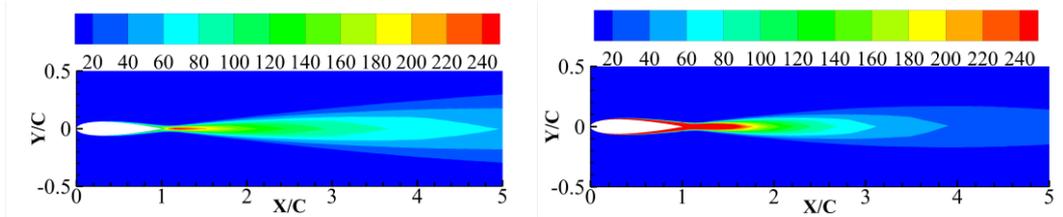

Figure 4 Left: contour the original eddy viscosity; right: contour of the eddy viscosity after transforming

Table 2 The initial input features of the model, where U and V are the velocity in the x,y direction, P is the pressure, ρ is the density, γ=1.4 is a constant, Y is the y coordinate, *dis* is the minimum distance from the grid node to the wall, and *sig* is the symbol Function, *tanh* is the hyperbolic tangent function. $q_{11}$ is an empirical feature characterizing the spatial properties of the flow field

| # | Description | Formula | # | Description | Formula |
|---|---|---|---|---|---|
| $q_1$ | X component of velocity | $U$ | $q_2$ | Norm of vorticity | $\sqrt{(\frac{\partial U}{\partial y}-\frac{\partial V}{\partial x})^2}$ |
| $q_3$ | Entropy | $\frac{\gamma P}{\rho^\gamma} - 1$ | $q_4$ | Direction of velocity | $\arctan(\frac{V sig(Y)}{U})$ |
| $q_5$ | Streamewise pressure gradient | $U_i \frac{\partial P}{\partial x_i}$ | $q_6$ | Ratio of pressure normal stresses to shear stresses | $\sqrt{\frac{\partial P}{\partial x_i} \frac{\partial P}{\partial x_i}}$ |
| $q_7$ | Strain rate | $(2S_{ij}S_{ij})^{1/2}$ | $q_8$ | Empirical function of dis | $dis^2 q_2(1 - \tanh(dis))$ |
| $q_9$ | Non-orthogonality between velocity and its gradient | $\left\| U_i U_j \frac{\partial U_i}{\partial x_j} \right\|$ | $q_{10}$ | Velocity projection | $sig(Y)[-V + U * \tan(\alpha)]$ |
| $q_{11}$ | Empirical function | $\exp\left(\sqrt{\frac{Dref1}{\min(dis)}}\right)\sqrt{\frac{Dref_0}{Dref_2}} - 2$, $Dref_0 = \frac{1}{\sqrt{Re}}$, $Dref_1 = \min(dis, Dref_0)$, $Dref_2 = \max(dis, Dref_0)$ | | | |

Besides input features, the output feature is also built to highlight the importance of the eddy viscosity in the boundary layer and reduce the difference under different Reynolds numbers. To this end, the following two measures have been taken:

(1) Introduce the weight on distance to change the distribution of eddy viscosity. The specific transformation formula is:

$$\hat{\mu}_t = \frac{\mu_t}{dis^p} \tag{6}$$

The contour of eddy viscosity after the transformation is shown in Figure 4. The value of eddy viscosity near the near-wall surface becomes larger after the transformation, thus reflecting the higher weight of the near-wall region.

(2) The eddy viscosity coefficient is divided by the corresponding Reynolds number, which makes eddy viscosity coefficients at different Reynolds numbers more similar. The comparison before and after the change was shown in Fig. 5. The distribution of eddy viscosity coefficient after dividing by Re was more similar, which is conducive to build the model.

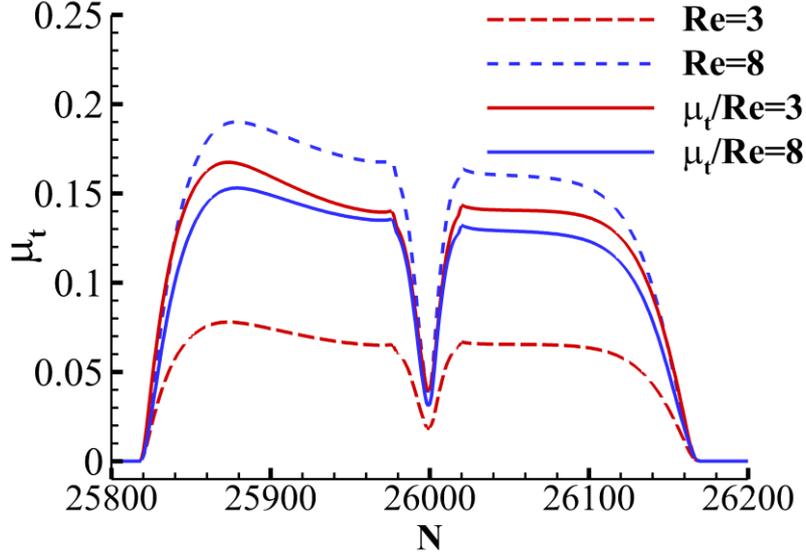

Figure 5 The contrast before and after dividing the eddy viscosity coefficient by Re.

In addition, we normalize the input and output features in order to eliminate the difference in magnitude between features:

$$\hat{x} = \frac{x - x_{\min}}{x_{\max} - x_{\min}} \tag{7}$$

where $x$ denotes a feature, $x_{max}$ denotes the maximum value of the feature, and $x_{min}$ denotes the minimum value of the feature. $\hat{x}$ is the normalized feature. After normalization, the magnitude of each feature is between 0 and 1.

To reduce the learning difficulty and improve the stability of the model, we evaluate the importance of the input features and deleted several features. There is still no accepted solution for how to select features reasonably. Here we use a method based on the importance of features to judge the value of each feature. The flow of the whole method is shown in Figure 6.

We use two ensemble machine learning models, Random Forest[40] and Lightgbm[41] to get the relative importance of features. Random Forest is a classic ensemble learning model. It is simple, easy to implement, has low computational overhead, and exhibits powerful performance in many tasks. Lightgbm is a framework that implements the GBDT (Gradient Boosting Decision Tree) algorithm. It supports high-efficiency parallel training, and has the advantages of faster training speed, lower memory consumption, better accuracy, support for distributed, and fast processing of massive data. The sub-models of both models are decision trees. At each step of the modeling process, the data set is divided according to a certain feature. In the CART decision tree algorithm, the Gini index is used to select the division attributes. The definition of the Gini index is as follows:

$$Gini(D) = \sum_{k=1}^{|y|} \sum_{k' \neq k} p_k p_{k'} = 1 - \sum_{k=1}^{|y|} p_k^2 \qquad (8)$$

Where $p_k$ is the proportion of the kth category samples in the sample set. The Gini index reflects the probability that two samples randomly selected from dataset D have inconsistent category markers. Therefore, the smaller the Gini index, the higher the purity of the dataset D. Because the two ensemble models (random Forest, Lightgbm) use different ensemble methods (bagging, boosting), it is considered that the two models give the importance of features from different perspectives. After comprehensively considering the importance of the two features, the least important features considered by the two models will be removed. If the fluctuation of the model`s performance before and after the removal is within the acceptable range, the feature is determined to be removed. If the model's performance deteriorates, the feature is no longer removed. The specific results of feature selection are shown in Section 3.1.

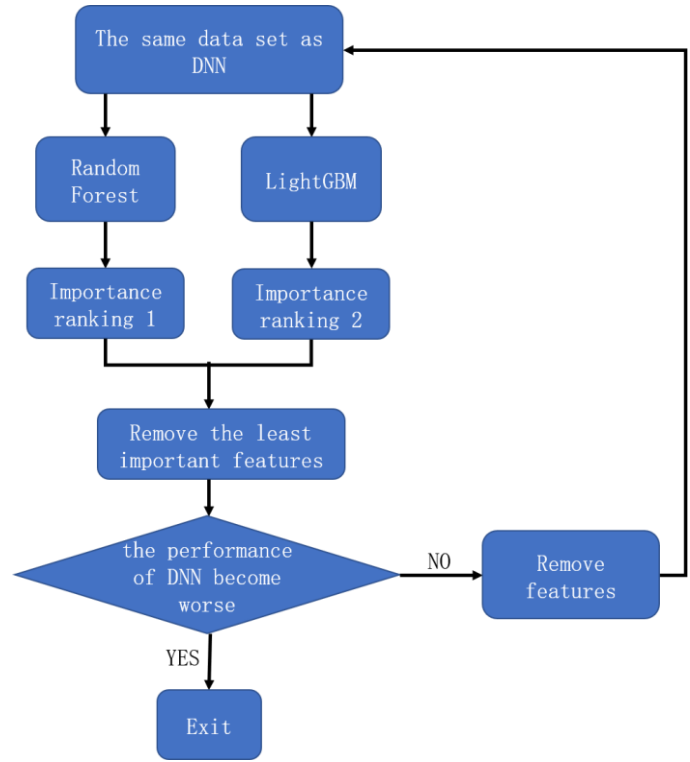

Figure 6 Feature selection method based on feature importance

## 2.5 Fully connected neural network

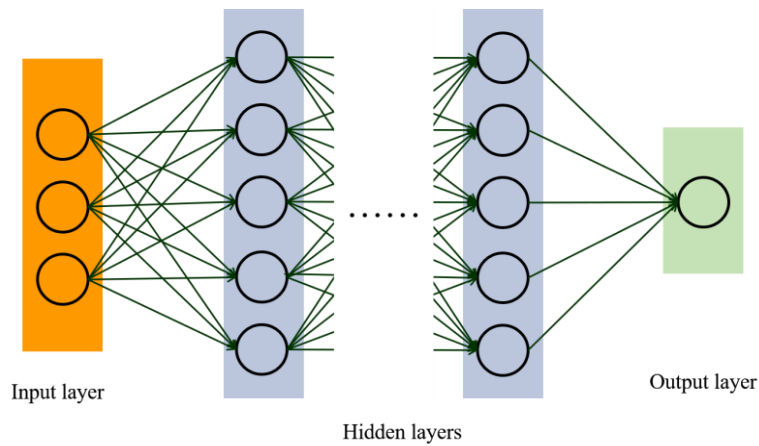

Figure 7 Schematic diagram of fully connected neural network

Fully connected neural network, also known as Multilayer Perceptron (MLP), is one of the most common neural network models. The schematic diagram of it is shown in Figure 7. The leftmost layer is the input layer, which is responsible for receiving data and passing it to the next layer. The middle layers are called the hidden layers, and the rightmost layer is called the output layer. The hidden layers and the output layer are responsible for processing and outputting data. The number of layers is the depth of neural network, and the number of neurons in the hidden layer is the width of neural network.

The connections between neurons in each layer have connection weights. And there are biases on each neuron in the hidden layer and the output layer. Learning process of the neural network is to adjust these weights and biases through the training data, so that the model can fit the data well. Therefore, it can be considered that what a neural network has learned is implicit in these weights and biases. Let the data received by the input layer be X, then the data passed to the hidden layer after the transformation is σ(WX+b), where W is the weight matrix and b is the bias vector. σ is the activation function, which introduces nonlinearity in the model. The commonly used activation functions are sigmoid, tanh, ReLU, etc. Their expressions are as follows:

$$sigmod(x) = \frac{1}{1+e^{-x}}$$
$$\tanh(x) = \frac{e^x - e^{-x}}{e^x + e^{-x}} \tag{9}$$
$$ReLU(x) = \begin{cases} x, & x > 0 \\ 0, & x \leq 0 \end{cases}$$

When modeling with neural networks, it is very important to determine the structural parameters (width, depth) and training parameters (learning rate, etc.) of the neural network. Although neural network structural search technology[42] and Bayesian optimization[43] methods are available, these methods often require more sophisticated technical support and higher computational resources, and there is uncertainty in the results. Therefore, here we still determine the values of each parameter based on experience and testing on a small data set. The neural network we use in this paper contains four hidden layers, each with 128,64,64,64 neurons. The number of neurons in the input layer is equal to the number of input features, while the output layer has only one neuron. The activation function between hidden layers is LeakyReLU[44].

$$LeakReLU(x) = \begin{cases} x, & x > 0 \\ \lambda x, & x < 0 \end{cases} \tag{10}$$

Where λ is an adjustable hyperparameter, the proposer of this function suggests taking it as 0.01 or even smaller, which is set to 0.2 in this paper by testing on the dataset. The activation of the hidden layer to the output layer is linear activation, i.e., $y = x$.

Loss function is another important factor to consider in the neural network modeling. When using neural networks for regression tasks, the commonly used loss functions are mean absolute error (MAE) or L1 loss, mean square error (MSE) or L2 loss as follows:

$$L_1(y, \hat{y}) = \frac{1}{N} \sum_{i=1}^{N} |y_i - \hat{y}_i| \tag{11}$$

$$L_2(y, \hat{y}) = \frac{1}{N} \sum_{i=1}^{N} (y - \hat{y})^2 \tag{12}$$

where $y$ is the true value and $\hat{y}$ is the predicted value. These two loss functions only consider the difference between the predicted values and sample values, which is a purely data-based perspective. When modeling physical problems, it is not sufficient to consider only the errors in the data. The predicted values of the model have to satisfy certain physical constraints while being accurate. Therefore, we design a loss function of the following form:

$$\begin{aligned} Loss(y, \hat{y}) = & a_1 \frac{1}{n} \sum_{i=1}^{n} (y - \hat{y})^2 + \\ & a_2 \frac{1}{n} \sum_{i=1}^{n} \max(0 - \hat{y}, 0) + \\ & a_3 \frac{1}{n} \sum_{i=1}^{n} \max(\hat{y} - 1, 0) + \\ & a_4 \frac{1}{n} \sum_{i=1}^{n} \max(y - \hat{y}, \beta y) + \\ & a_5 \frac{1}{n} \sum_{i=1}^{n} (\tau_i - \hat{\tau}_i)^2 \end{aligned} \tag{13}$$

where $a_i (i = 1, 2, 3, 4)$ is the proportion of the error in each component, which can be adjusted according to the actual situation.

In Equation (13), the first term is the MSE loss of data. The second term represents the loss arising from the fraction of model predictions less than 0. Here, a priori knowledge is that the eddy viscosity will not be a negative value. Therefore, it is necessary to put some constraints on the negative values in the model predictions. When the prediction is negative, the second term generates losses to guide the model to learn in a more physically consistent direction. The third term has a similar meaning to the second term and serves to constrain the model to produce predictions greater than one. This is mainly because we normalize the data to the range between 0 and 1. The fourth term is a relative error between the predicted and true values, which is constructed based on the a priori that "the eddy viscosity coefficient inside the boundary layer is small but more important". β is an adjustable constant that indicates the magnitude of the tolerable relative error. The last error is the error about Reynolds stress, where the stress term is calculated according to the following formula.

$$\tau = 2\mu_t \|S_{ij}\|$$
$$S_{ij} = \frac{1}{2}(\frac{\partial U_i}{\partial x_j} + \frac{\partial U_j}{\partial x_i}) \qquad (14)$$

where $\|S_{ij}\|$ is the norm of the strain rate tensor $S_{ij}$. This loss is also added to highlight the importance of the eddy viscosity inside the boundary layer and to constrain the network learning process from the perspective of Reynolds stress. This loss also ensures that the coupled iterative solution proceeds smoothly to a certain extent since it is not the eddy viscosity coefficient but the stress term that is involved in the coupled calculation.

By adjusting the coefficients in front of each loss term, the model can be targeted to learn in a certain direction. We set it to 0.6, 0.1, 0.1, 0.1, 0.1, respectively. In a specific training process, the above five losses are at the same magnitude, and they decrease consistently as the training progresses.

The neural network model used in this paper is based on Pytorch[45]. The optimization algorithm for training is the Adam algorithm[46] with batch size of 128 and an initial learning size of 0.003. As the training continues and the error decreases, we use a dynamic adjustment to reduce the learning rate. The specific method is to change the learning rate to the original 0.5 when the model's loss on the validation set no longer decreases. Otherwise, training epoch is 300. When the performance of the model on the training set and the validation set tends to be stable, the error is on the order of $10^{-5}$.

## 2.6 Coupling method

Constructing a model that maps the time-averaged flow features to the eddy viscosity is only a stage in our work. We eventually want to replace the traditional turbulence model (here the SA model) with the resulting model and implement a mutual coupling with the NS equations.

As shown in Figure 1, in each iteration of the coupling, the CFD solver passes the corresponding input features to the neural network model, and the neural network outputs the corresponding eddy viscosity to the CFD solver. Here the neural network model and the CFD solver influence each other. We realize the data transfer process between Fortran and Python program with the help of CFFI module in Python.

We use the spatially smooth method to improve the convergence and stability of the coupling process. The spatial smoothing method uses the weighted average of the eddy viscosity on the current grid and all its neighboring grids as the final eddy viscosity on the current grid. This method can make the distribution of eddy viscosity on the space smoother and reduce the generation of singular values.

## 3. Results

### 3.1 Results of feature selection

The feature selection method based on feature importance proposed in section 2.3 is used to perform feature selection on the initial 11 features. The feature importance obtained by the random forest and Lightgbm model is shown in Figure 8. According to the importance, we can see that there are some features that are low in the importance given by both models. These are the features that need to be focused on when making feature selection.

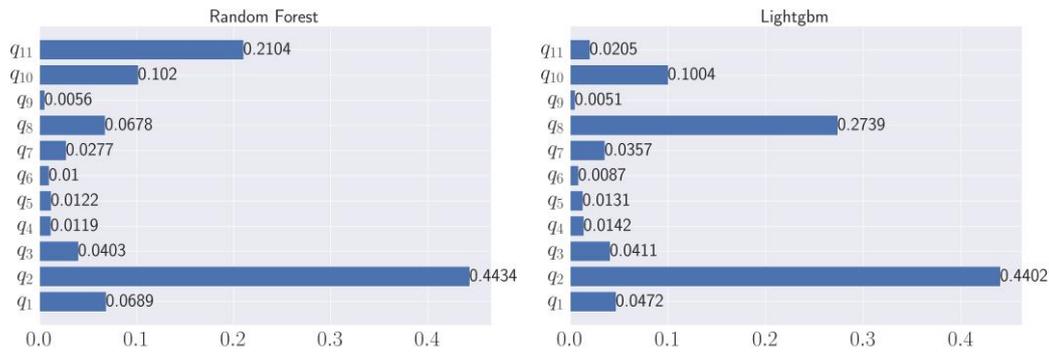

Figure 8 Feature importance given by Random Forest(left) and Lightgbm(right)

After using the feature importance-based method for feature selection, three features are finally removed: $q_5$, $q_6$, and $q_9$. It is worth noting that Yang et al.[47] also evaluate the feature importance using the random forest model. In their work, flow pressure gradient was found to be the most important feature($q_5$). This feature is removed in this paper. This shows that the importance of the features given by the random forest is different and worthy of reference in different problems. After removing these three features, a comparison of the loss curves in the DNN modeling is shown in Figure 9.

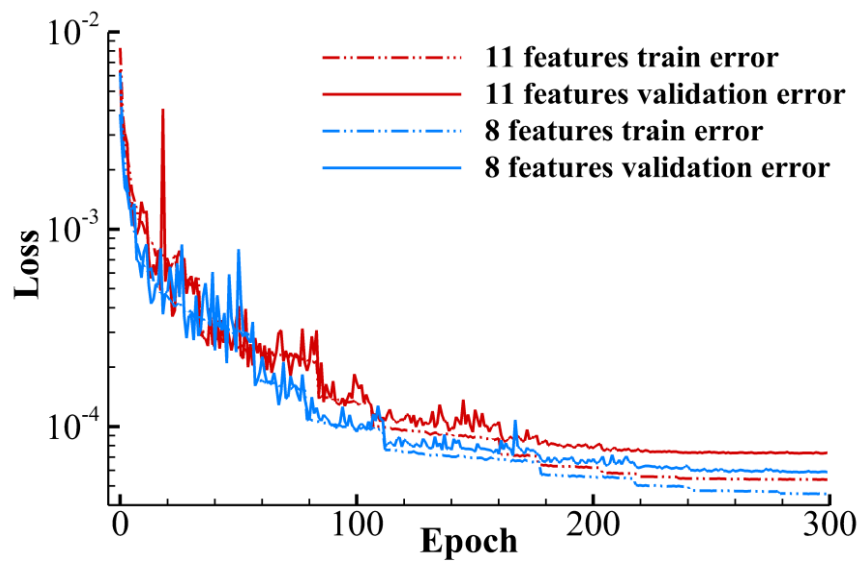

Figure 9 Loss curve of DNN based on 8 features and 11 features

The comparison of the loss curves shows that the loss of the model with 11 features and with 8 features eventually converge to a similar magnitude. The training error based on 8 features is even lower than that based on 11 features, and the same is true for the validation error. This shows the reasonableness of our feature selection method. Through feature selection, the input dimension of the model decreases, and the complexity and learning difficulty of the model are somewhat alleviated

The benefits of feature selection are not limited to model training. After reducing redundant features, the stability and robustness of the model are also improved. This is mainly reflected in the process of mutual coupling with CFD solver. The model based on 8 input features shows better stability and convergence in the coupling process than the model based on 11 input features. And the residual convergence in transonic states is even better than the original SA model.

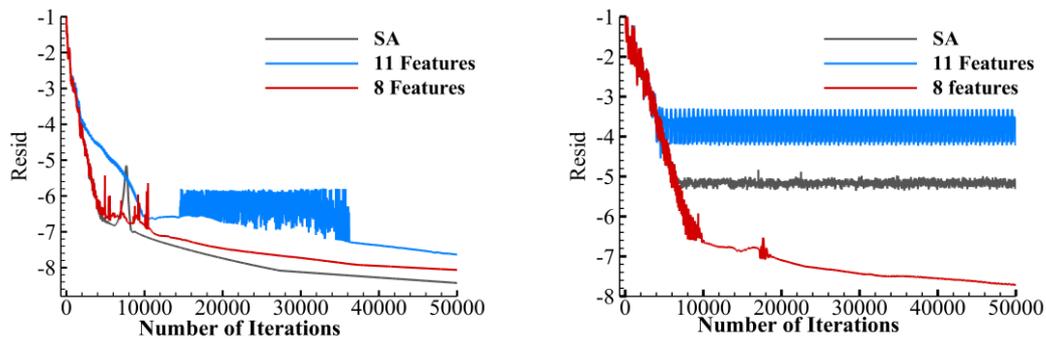

Figure 10 Residual convergence curve. Left: Ma=0.4,α=5.07,Re=4e6   Right: Ma=0.7,α=2.48,Re=5e6

The left figure in Figure 10 shows that the model based on 11 input features appears unstable during the iterative process, while the residual curve of the model based on 8 input features does not show significant fluctuations and the final convergence result is better than the former. This reflects that the model with 8 input features has better stability and convergence properties. The right figure in Figure 10 shows that the model based on 11 input features goes directly into a large oscillation at the transonic velocity, while the SA model also shows a small oscillation. The model based on 8 input features, on the other hand, shows a stronger stability, arriving at a much smaller residual in the same number of iteration steps.

## 3.2 Results of neural network modeling

The computation mesh adopts a hybrid mesh. The overall mesh and the local mesh of the leading edge are shown in Figure 11. The model has good generalization for state variables such as Mach number, angle of attack, Reynolds number and airfoils. The eddy viscosity predicted by the model under different conditions are given below.

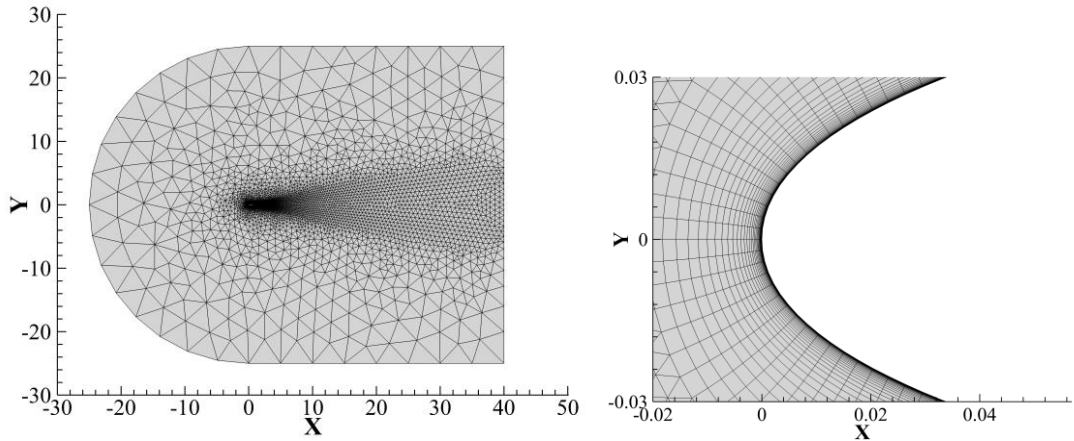

Figure 11 Schematic diagram of computational mesh

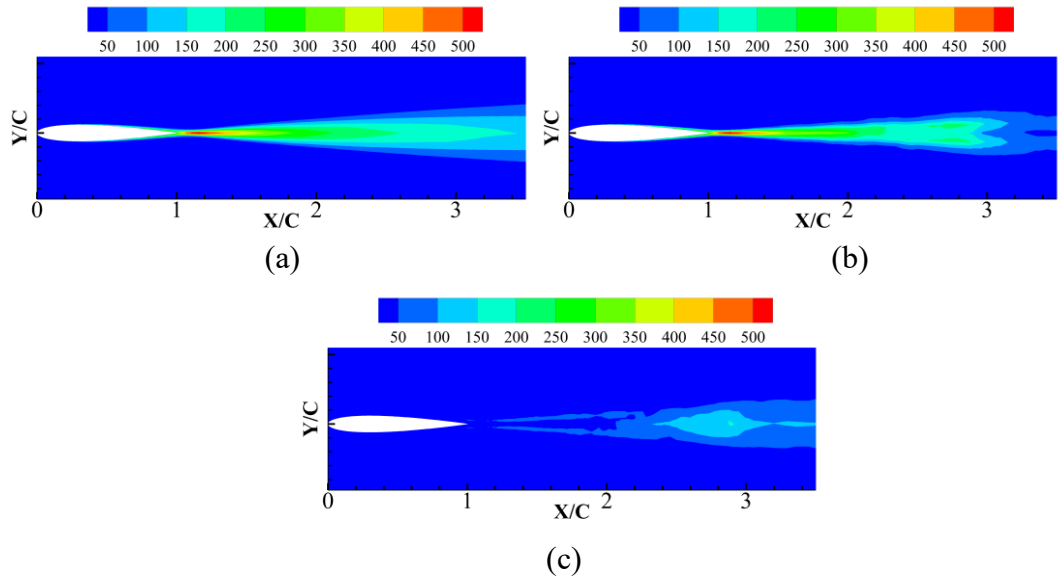

Figure 12 Training case Ma=0.1, α=0 deg, Re=9e6 The contour of eddy viscosity computed by (a) SA model (b) DNN model and (c) the absolute error contour

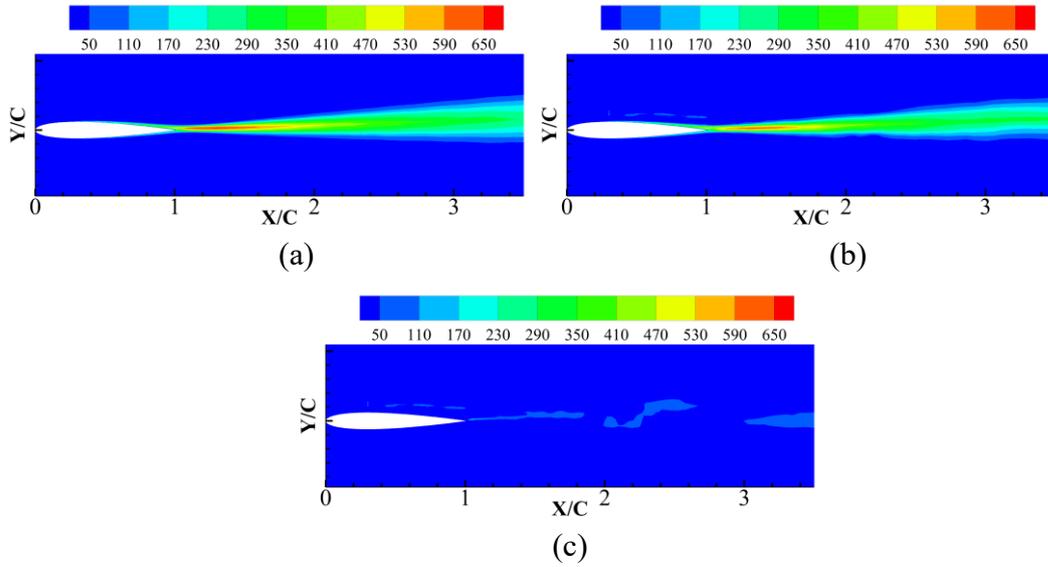

Figure 13 Testing case Ma=0.71, α=2.48 deg, Re=5e6   The contour of eddy viscosity computed by (a) SA model (b) DNN model and (c) the absolute error contour

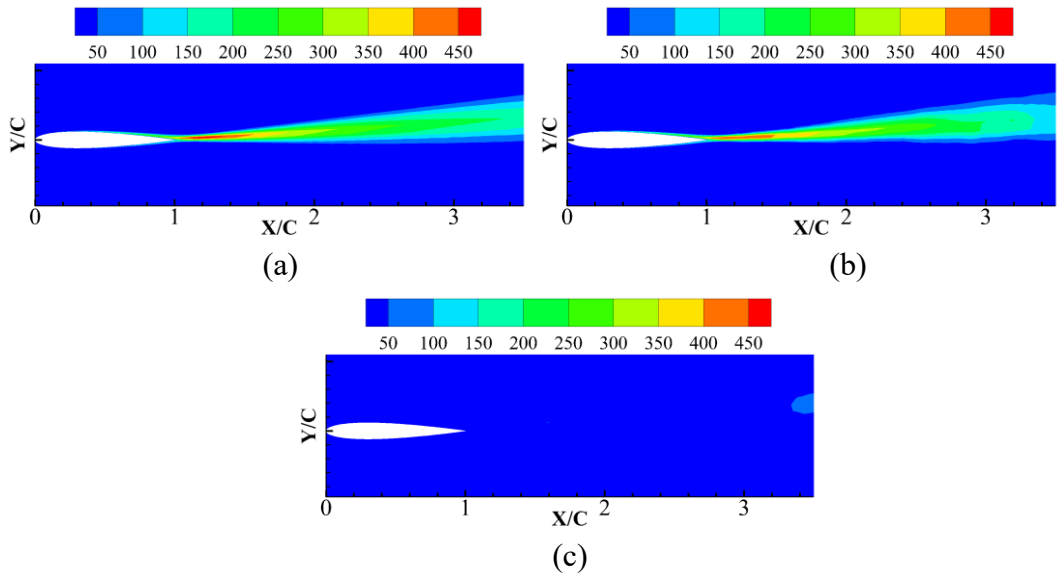

Figure 14 Testing case Ma=0.4, α=5.07 deg, Re=4e6 The contour of eddy viscosity calculated by (a) SA model (b) DNN model and (c) the absolute error countour

The predicted values of the DNN model are close to the real values. And the absolute errors between them are mainly concentrated in the wake region, which is mainly because the attention to the eddy viscosity in the wake region is not very high in the modeling process. Moreover, the errors in this part do not have much influence on the overall velocity distribution.

To demonstrate the predictive power of the model more clearly, we also calculate the regression coefficient of determination R2 between the predicted and true values. A linear fit to the predicted values is also performed with the true values as independent variables. Ideally, the linear fit function between the predicted and true values should be a proportional function. The deviation of the predicted and true values at different magnitudes can be seen by comparing the true fit function with the linear function, as shown in Figure 15.

$$R^2 = 1 - \frac{\sum_{i=1}^{n}(\hat{y}_i - y_i)^2}{\sum_{i=1}^{n}(y_i - \overline{y})^2} \tag{15}$$

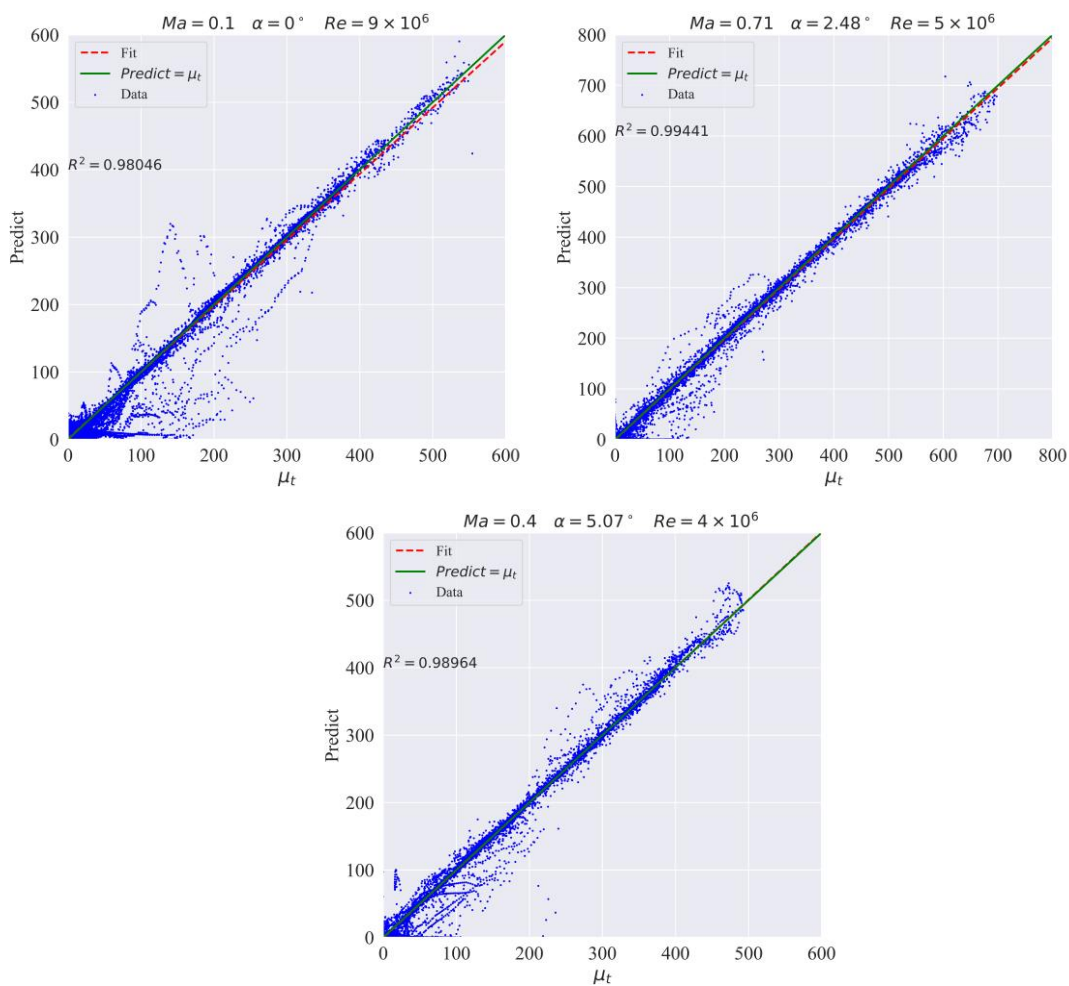

Figure 15 An a priori statistical assessment of the trained DNN model

## 3.3 Results of mutual coupling

The coupling results under different test cases are shown below, respectively.

（1）Testing case $Ma=0.4, \alpha=5.07°, Re=4\times10^6$

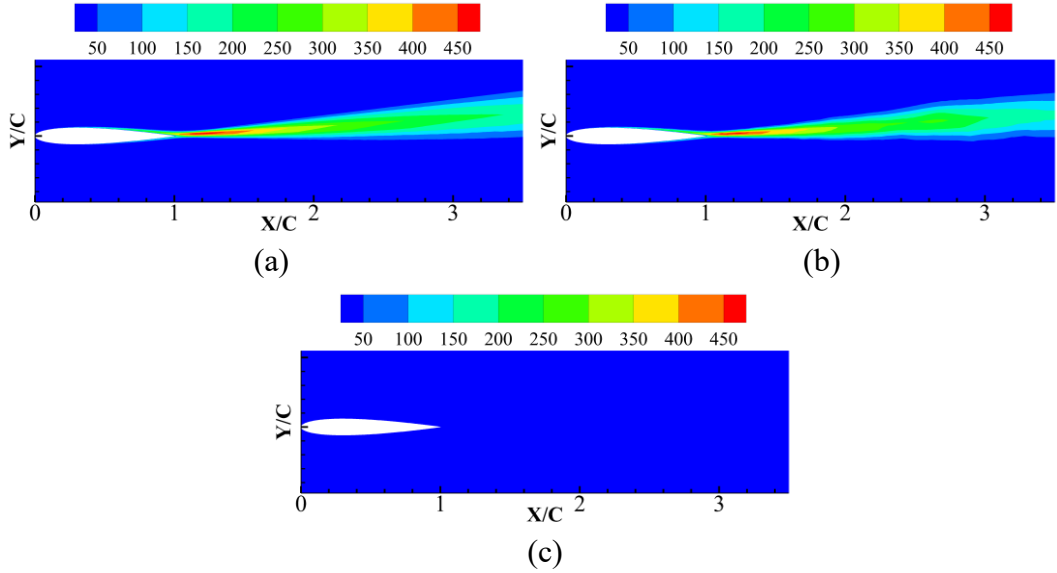

Figure 16 (a) the eddy viscosity contour computed by the SA model; (b) the final eddy viscosity contour obtained by coupling the DNN model with the CFD solver; (c) the absolute error between (a) and (b)

Comparing the results in Figure 14 with those in Figure 16 , Figure 14 shows only the difference between the predicted values and the real values, while Figure 16 shows the stable eddy viscosity obtained when the neural network model is coupled with the CFD solver to the convergence state. The results of the distribution of the eddy viscosity along the normal distribution are shown in Figure 17. The pressure and friction coefficients are distributed as follows (Figure 18).

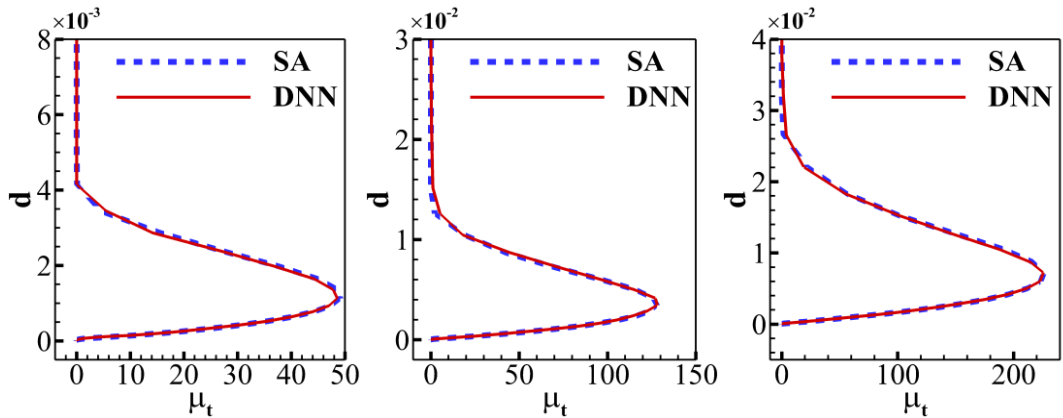

Figure 17 Eddy viscosity normal distribution at different positions of airfoil

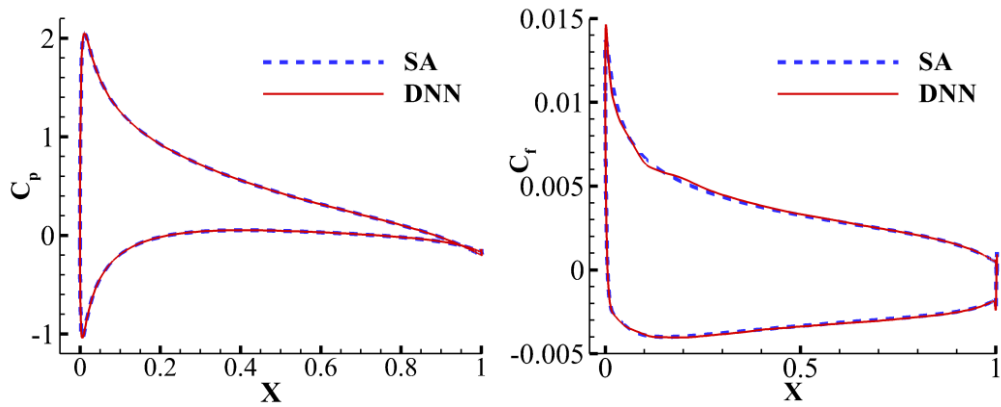

Figure 18 Left: Pressure coefficient. Right: Skin-friction coefficient

（2）Testing case $Ma = 0.71, \alpha = 2.48°, Re=5\times10^6$

Similar to the above test case, the comparison of the obtained eddy viscosity contour is shown in Figure 19 , the normal distribution of vortex viscosity coefficient is shown in Fig. 20, and the distribution of pressure and friction resistance is shown in Fig. 21.

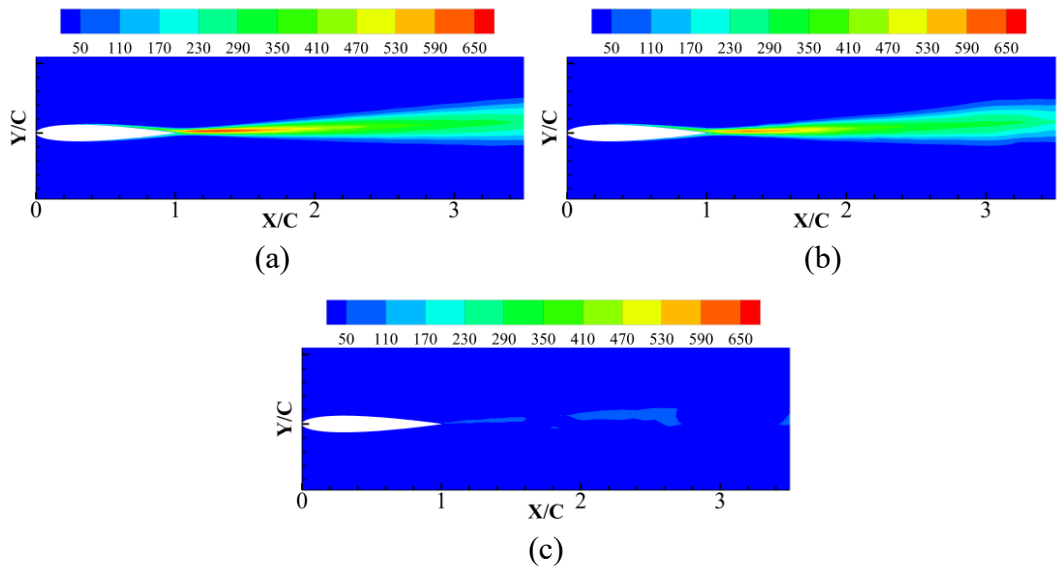

Figure 19 (a) the eddy viscosity contour computed by the SA model; (b) the final eddy viscosity contour obtained by coupling the DNN model with the CFD solver; (c) the absolute error between (a) and (b)

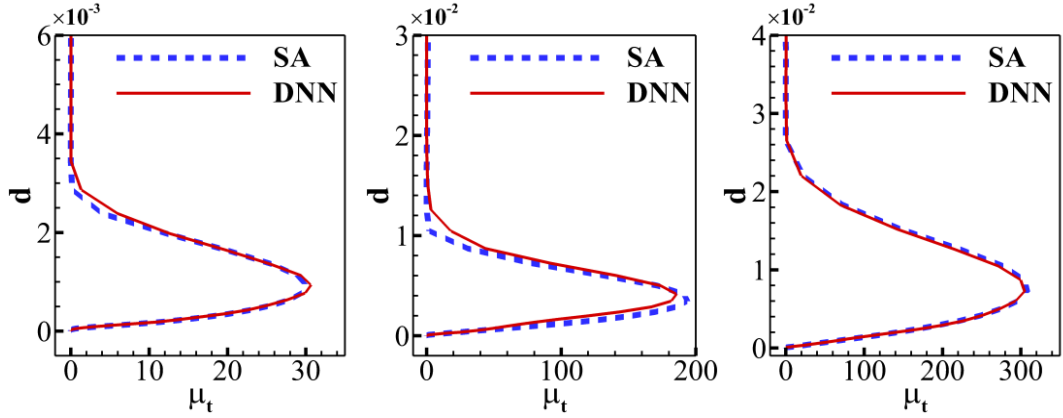
Figure 20 Eddy viscosity normal distribution at different positions of airfoil

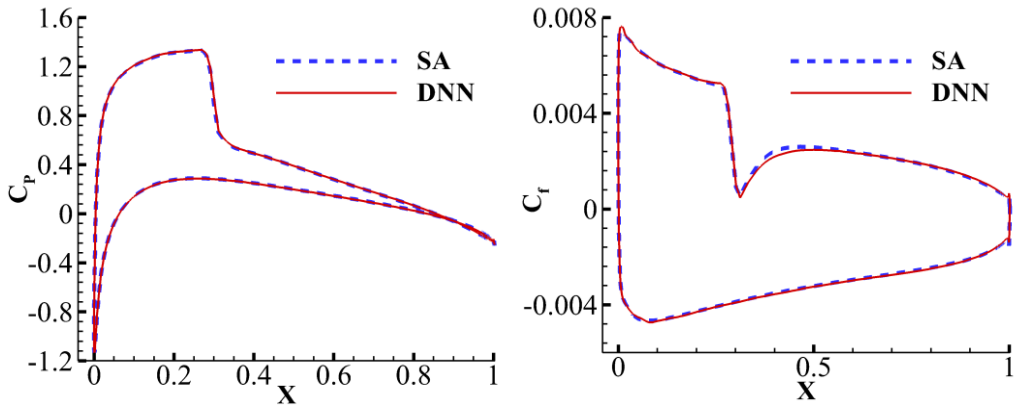
Figure 21 Left: Pressure coefficient. Right: Skin-friction coefficient

Both two test states above show that the DNN model can make reasonable predictions of the eddy viscosity during the coupled iterations. The final force coefficient distributions are also in good agreement with the results of the SA model.

(3) Testing case on NACA0014  $Ma = 0.5, \alpha = 5.0°, \mathrm{Re}=3\times10^6$

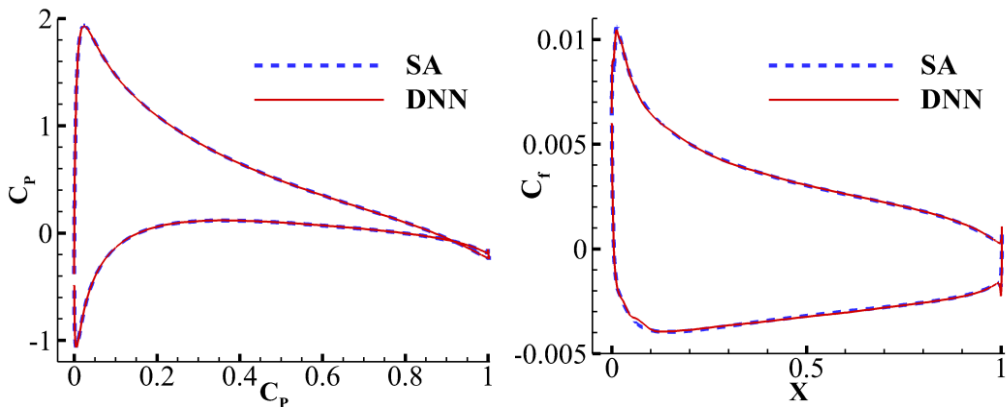
Figure 22 Left: Pressure coefficient. Right: Skin-friction coefficient

(4) Testing case on RAE2822 airfoil $Ma = 0.23, \alpha = 2.95°, \text{Re}=3\times10^6$

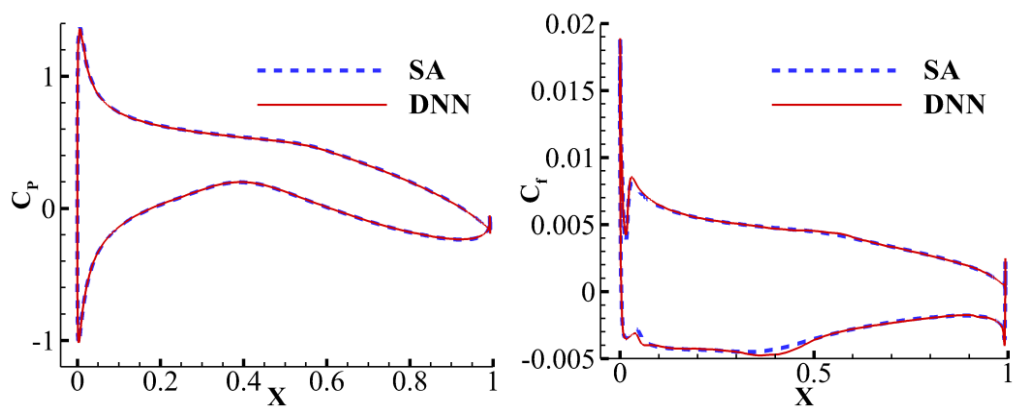

Figure 23 Figure 24 Left: Pressure coefficient. Right: Skin-friction coefficient

The above two test cases show that the model also has a certain generalization on different airfoils.

To further test the capability of the data-driven turbulence model, a series of numerical experiments were conducted. The statistics of the results are shown in Table 3. From the result, we can see that the data-driven turbulence model performs very close to the SA model.

Table 3 The lift/drag coefficients obtained by the two models under different flow conditions

| Ma | α/° | Re/$10^6$ | $C_L$ (SA/DNN) | | $C_D$ (SA/DNN) | | relative error ($C_L$ $C_D$) | |
|---|---|---|---|---|---|---|---|---|
| 0.24 | 9.35 | 3 | 1.007 | 1.008 | 0.015 | 0.0149 | 0.0993% | 0.6667% |
| 0.15 | 0.75 | 3 | 0.082 | 0.082 | 0.009 | 0.009 | 0 | 0 |
| 0.226 | 2.95 | 3 | 0.327 | 0.326 | 0.01 | 0.01 | 0.3058% | 0 |
| 0.174 | 7.53 | 3 | 0.81 | 0.807 | 0.012 | 0.013 | 0.3704% | 8.3333% |
| 0.293 | 6.42 | 3 | 0.716 | 0.715 | 0.012 | 0.012 | 0.1397% | 0 |
| 0.215 | 8.25 | 3 | 0.892 | 0.898 | 0.013 | 0.014 | 0.6726% | 7.6923% |
| 0.13 | 2.43 | 3 | 0.264 | 0.264 | 0.009 | 0.009 | 0 | 0 |
| 0.71 | 2.48 | 5 | 0.406 | 0.399 | 0.013 | 0.013 | 1.754% | 0 |
| 0.4 | 5.07 | 4 | 0.593 | 0.044 | 0.594 | 0.044 | 0.168% | 0 |
| 0.35 | 6.37 | 6 | 0.727 | 0.728 | 0.0109 | 0.0110 | 0.137% | 0.9% |
| 0.27 | 11.41 | 7 | 0.123 | 0.124 | 0.0165 | 0.0157 | 0.813% | 4.84% |
| 0.17 | 10.15 | 9 | 0.109 | 0.108 | 0.0128 | 0.0134 | 0.9% | 4.69% |

## 4. Conclusion & Outlook

In this paper, a turbulence model is constructed using a fully connected neural network based on the NACA0012 airfoil turbulence data simulated by the SA model. The accuracy and generalization ability of the data-driven turbulence model for different flow states and different airfoil types are verified by comparing the results with those of the SA model.

In the specific modeling process, we implement a feature selection method based on feature importance, and successfully applied it to the input features. The results show that this feature selection method can effectively remove redundant features. The model constructed based on the new input features has better accuracy and stability in mutual coupling with the CFD solver. Furthermore, we have also designed a new loss function, which is different from the previous loss function purely based on data. The new loss function contains physical information, which makes the predicted value of the model more in line with physics. The introduction of these terms also helps to reduce the overfitting tendency and improve the generalization ability of the model. The frictional and pressure coefficient distribution computed by the coupling of the DNN turbulence model and the CFD solver are very close to the results of the SA model. And the stability of the new model is better than those of the original model under partial transonic conditions.

Our ultimate goal is to obtain more reliable results in cases where the traditional RANS turbulence model cannot provide accurate results, which requires high confidence flow data for modeling. However, in the case of high Reynolds numbers, data from DNS/LES are hard to obtain. Therefore, we first selected some simple examples to verify the feasibility of the data-driven modeling approach. This is a preliminary exploration for future work. Encouraged by this work, we hope to build better data-driven turbulence models than traditional models when high-fidelity data are generated by data assimilation or other methods. So far, the computational convergence of the data-driven model is better than the traditional model in some conditions. The solution efficiency is also higher than that of the traditional model, and the single computation time is only 40% of that of the traditional model.

**ACKNOWLEDGEMENTS**

This paper was supported by the National Natural Science Foundation of China (no. 91852115) and funded by Beijing Key Laboratory of Civil Aircraft Design and Simulation Technology.